\documentclass[openacc]{rstransa}

\def\dd {{\rm d}}

\begin{document}

\title{The Serendipity of Electroweak Baryogenesis}

\author{
G\'eraldine Servant$^{1,2}$}

\address{$^{1}$DESY, Notkestra{\ss}e 85, D-22607 Hamburg, Germany\\
$^{2}$II. Institute of Theoretical Physics, University of Hamburg, D-22761 Hamburg}

\subject{particle physics, cosmology}


\corres{Insert corresponding author name\\
\email{geraldine.servant@desy.de}}

\begin{abstract}

The origin of the matter antimatter asymmetry of the universe remains unexplained in the Standard Model of particle physics. The origin of the flavour structure is another  major puzzle of the theory.
In this article, we report on recent work attempting to link the two themes through the appealing framework of electroweak baryogenesis. 
We show that Yukawa couplings of Standard Model fermions can be the source of CP-violation for electroweak baryogenesis 
if they vary at the same time as the Higgs is acquiring its vacuum expectation value, 
offering new avenues for electroweak baryogenesis. 
The advantage of this approach is that it circumvents
the usual severe bounds from Electric Dipole moments.
These ideas  apply if  the mechanism explaining
the flavour structure of the Standard Model  is connected to electroweak symmetry breaking, as motivated 
for instance in Randall-Sundrum or Composite Higgs models. 
We compute the resulting baryon asymmetry for different configurations  of the Yukawa coupling variation across the bubble wall, and show that it can naturally be of the right order.

\end{abstract}




\maketitle

\section{Introduction}

All fermion masses in the Standard Model, apart form the Top quark, are much smaller than the electroweak (EW) scale. Besides, the mass spectrum  ranges many orders of magnitude.
The mysterious structure of masses and mixing angles  is the so-called flavour puzzle. While many solutions have been proposed, the cosmological aspects  of the corresponding models have not been studied. On the other hand, in most cases, Yukawa couplings are dynamical and it is natural to investigate about their cosmological evolution. In this article we report on recent work connecting flavour cosmology and baryogenesis \cite{Baldes:2016rqn,Baldes:2016gaf,vonHarling:2016vhf,Bruggisser:2017lhc,Bruggisser:2018mus,Bruggisser:2018mrt,Baldes:2018nel}.

Electroweak baryogenesis (EWBG) is a mechanism to explain the matter antimatter asymmetry of the universe  using Standard Model baryon number violation \cite{Kuzmin:1985mm}. It relies on a charge transport mechanism in the vicinity of bubble walls during a first-order EW phase transition \cite{Cohen:1990it}. It is particularly attractive as it relies on EW scale physics only and is therefore testable experimentally. It requires an extension of the Higgs sector leading to a first-order EW phase transition. This is minimally achieved by adding a singlet scalar to the Standard Model. In EWBG, CP violation  comes into play when chiral fermions scatter off the Higgs at the phase interface. A chiral asymmetry is created in front of the bubble wall that is converted by sphalerons into a baryon number.  CP violation in the Standard Model is too suppressed to explain the baryon asymmetry \cite{Gavela:1993ts}.
New sources that have been commonly  studied in the literature are either in the chargino/neutralino mass matrix \cite{Carena:1997gx, Carena:2000id, Carena:2002ss,Konstandin:2005cd, Cirigliano:2009yd, Li:2008ez} or the sfermion sector in supersymmetric models  \cite{Carena:1997gx,Carena:2000id,Chung:2008aya,Kozaczuk:2012xv}  or coming from a varying top quark Yukawa coupling \cite{Fromme:2006wx,Bodeker:2004ws} as motivated in composite Higgs models \cite{Espinosa:2011eu} or  in Two-Higgs doublet models \cite{Cline:1995dg,Cline:1996mga,Fromme:2006cm,Cline:2011mm,Dorsch:2016nrg,Alanne:2016wtx} where the CP violating source comes from the changing phase in the Higgs VEV  during the EW phase transition. A new possibility was studied recently where CP violation is coming from the dark matter particle \cite{Cline:2017qpe}.
A typical constraint on EWBG comes from Electric Dipole Moments, which is particularly severe in the case of supersymmetric scenarios \cite{Cirigliano:2009yd}.


In this article,  we consider that the source of CP violation has changed with time, which is a natural way to evade constraints. Following the same philosophy, strong CP violation from the QCD axion was studied in the context of cold baryogenesis in \cite{Servant:2014bla}. 
We are now investigating the possibility that the structure of the Cabibbo-Kobayashi-Maskawa (CKM) matrix is varying during the EW phase transition such that Yukawa couplings start with natural values of order one in the EW symmetric phase and end up with their present values in the broken phase. This way, CP violation is no longer suppressed by small Yukawa couplings during the EW phase transition \cite{Berkooz:2004kx}.  The main motivation is to link EWBG to low-scale flavour models.  If the physics responsible for the structure of the Yukawa couplings is linked to EW symmetry breaking, we can expect the Yukawa couplings to vary at the same time as the Higgs is acquiring a VEV, in particular if the flavour structure is controlled by a new scalar field which couples to the Higgs.
In principle, one  has to compute the full scalar potential, to determine the field path in a multiple scalar field space and compute the evolution of the Yukawa coupling during the phase transition. 
Particular realisations in explicit constructions were studied in  \cite{Baldes:2016gaf,vonHarling:2016vhf,Bruggisser:2018mus,Bruggisser:2018mrt,Baldes:2018nel}.
It is crucial to show that the Yukawa can vary with the Higgs, i.e. that the Higgs and Flavon fields vary simultaneously and not subsequently. 
For this to happen, the new scalar has to be light, i.e. not much heavier than the Higgs.
This is typically in tension with flavour bounds in Froggatt-Nielsen constructions \cite{Baldes:2016gaf}. However, in Randall-Sundrum and Composite Higgs models, this can be achieved \cite{vonHarling:2016vhf,Bruggisser:2018mus,Bruggisser:2018mrt}. 

We stress that in this framework, the Yukawa couplings do not depend explicitly on the Higgs VEV. Instead, the Yukawas are controlled by some other scalar field. However, because of the couplings between this additional scalar and the Higgs field, the Yukawa coupling effectively varies at the same time as the Higgs during the EW phase transition. A dependence of the Yukawa coupling on the Higgs VEV is induced during the EW phase transition, but today there is no such dependence.  Our scenario is therefore very different from \cite{Bauer:2015fxa,Bauer:2015kzy}, which is very constrained experimentally.

\section{Calculation of the baryon asymmetry}

\begin{figure}[t]
\centering\includegraphics[width=4.5in]{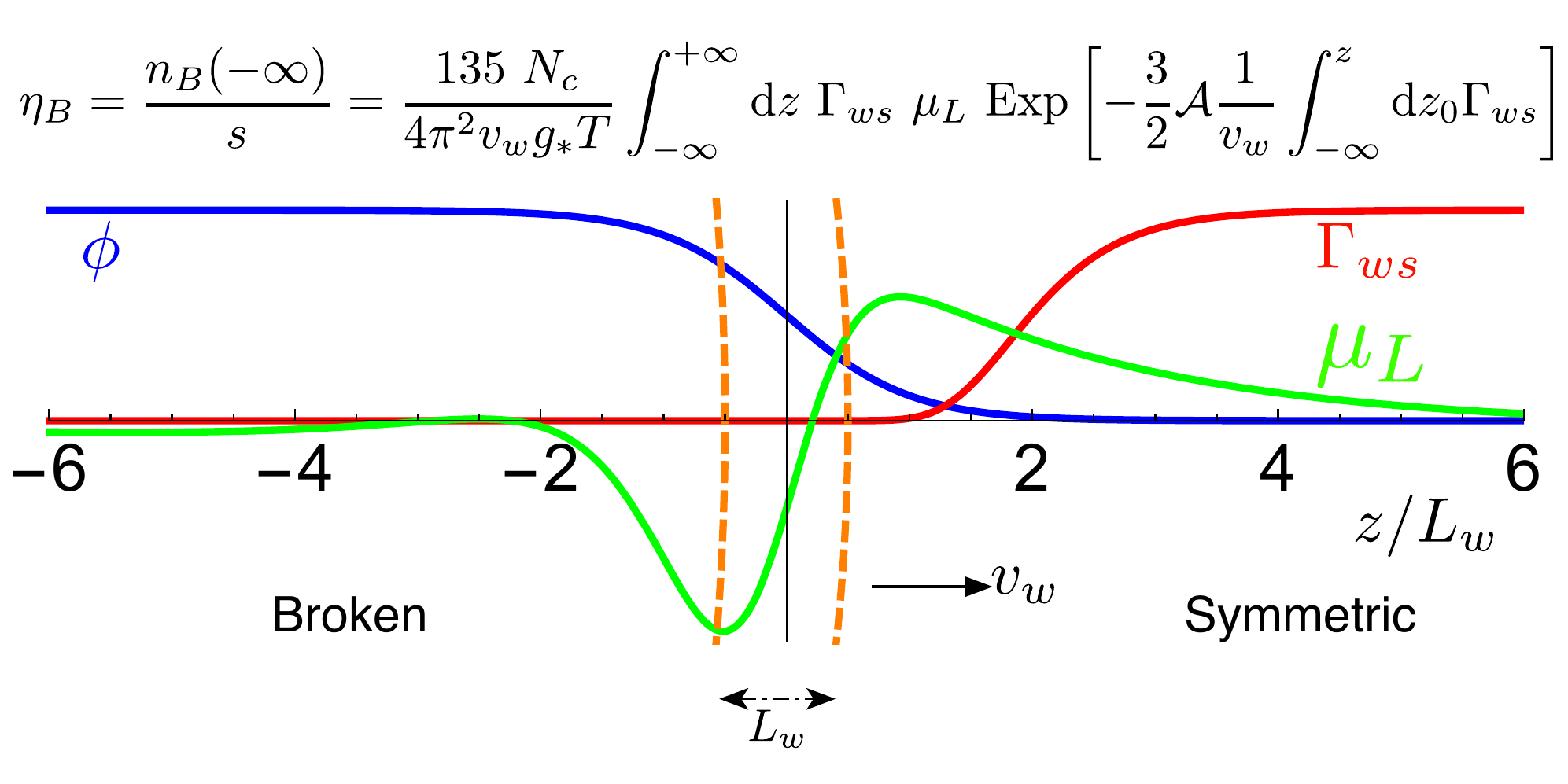}
\caption{A cut through the bubble wall, which moves from the left to the right (in the direction of positive $z$, i.e. $v_w>0$). In blue we show the profile of the Higgs VEV through the bubble wall. The rate for the sphaleron transitions (red) becomes important only in front of the bubble wall. The green line is the profile of the total chemical potential for the quark doublets. All curves are in arbitrary units, for illustration only. All the plotted quantities enter the calculation of the final baryon asymmetry of the universe $\eta_B$. Figure is taken from \cite{Bruggisser:2017lhc}.}
\label{fig:SchematicBubbleWall}
\end{figure}

During a first-order EW phase transition, bubbles are created, symmetric and broken phases coexist until bubbles percolate and the whole universe is converted into the broken phase.
To compute the  baryon asymmetry produced during bubble expansion, it is enough to focus on one single bubble and integrate the baryonic density over the radial coordinate perpendicular to the bubble wall. The bubble expands fast with a velocity $v_w$ through the plasma and we can approximate the bubble wall to  be planar with a characteristic thickness $L_w$.
Inside the bubble the Higgs has a non-vanishing vacuum expectation value $\phi$. Outside the bubble, the EW symmetry is unbroken $\phi=0$. 
We define the direction perpendicular to the bubble wall as $z$ and we choose the origin of our comoving (with the bubble wall) coordinate system as the middle of the wall. The phase where the EW symmetry is broken is for negative $z$ and the symmetric phase is for positive $z$. As the wall is passing, particles in the plasma get reflected and transmitted depending on their interactions with the wall. If some of these interactions are CP-violating, left- and right-handed particles will have different reflection and transmission coefficients and  a chiral asymmetry will develop inside and in front of the bubble wall. The resulting excess of left-handed fermions in front of the bubble wall can be converted into a net baryon number by the weak sphalerons, which are unsuppressed in the symmetric phase in front of the bubble. 
The whole mechanism is illustrated in figure~\ref{fig:SchematicBubbleWall}. 

The baryon asymmetry  $\eta_B$, defined as  the difference between the baryon and anti-baryon number densities normalised to the entropy density $s$ is given by: 
\begin{equation}\label{eqn:totalBaryonAsymmetry}
  \eta_B=\frac{n_B(-\infty)}{s}=\frac{135~N_c}{4\pi^2 v_w g_* T }\int_{-\infty}^{+\infty} \dd z ~ \Gamma_{ws}~\mu_L~ e^{-\frac{3}{2} {\cal A} \frac{1}{v_w}\int_{-\infty}^{z} \dd z_0 \Gamma_{ws}}\, ,
\end{equation}
where $\Gamma_{ws} = 10^{-6} \, T \, \exp(- a \phi(z)/T)$ is the $\phi$-dependent  weak sphaleron rate, $a \simeq 37$, 
$s=\frac{2\pi^2}{45}g_*T^3$, $g_*$ the number of relativistic degrees of freedom, ${\cal A}= 15 /2$, $N_c=3$ and $\mu_L$ is  the density of the excess of left-handed fermions in front of the bubble wall. The source term  
 $\mu_L$  depends on how fermions are transported through the bubble wall. This is determined by  solving a diffusion system of coupled differential equations, the so-called transport equations.
From dimensional analysis, the baryon asymmetry (\ref{eqn:totalBaryonAsymmetry}) will scale like 
\begin{equation}
\eta_B \sim \frac{\Gamma_{ws} \mu_L L_w}{g_* T} \sim \frac{10^{-8} \mu_L}{T} \mbox{ for } L_w\sim \frac{1}{T}
\end{equation}
so it is clear that for EWBG to be successful, $\mu_L/T$ has to be large and this requires new sources of CP violation beyond the Standard Model.

\section{New CP-violating sources from varying Yukawas}

The baryon asymmetry (\ref{eqn:totalBaryonAsymmetry}) can be written as:
\begin{equation}
  \eta_B= \sum_i \int_{-\infty}^{+ \infty}\dd y ~K_i(y) ~\bar{\cal S}_i(y)
\end{equation}
where we have separated the CP-violating source term $\bar{\cal S}_i(z)$ from the kernel $K_i(z)$ encoding effects of the diffusion through the bubble wall and the weak sphaleron. 
It was shown in \cite{Bruggisser:2017lhc} that at the lowest order in a derivative expansion 
of  Kadanoff-Baym equations that coincide with Boltzmann equations, 
 the CP-violating source is proportional to 
\begin{equation} 
  \Im\left[V^\dagger {m^\dagger}'' m V\right]_{ii} 
\label{eq:CPsource}
\end{equation}
where $V$ is the transformation matrix which diagonalizes $m^{\dagger}m$ and $m$ is the fermion mass matrix. The subscript $ii$ refers to the diagonal entries and does not stand for the conventional summation.
It is then easy to see that the CP-violating part vanishes for the Standard Model which has constant Yukawa couplings. Indeed, for constant Yukawas $\Im\left[V^\dagger{m^\dagger}''mV\right]\propto\Im\left[V^\dagger{Y^\dagger}YV\right]\phi''\phi$ and since  $V^\dagger{Y^\dagger}YV$ is hermitian, the diagonal entries are real.

\begin{figure}[t]
\centering
\includegraphics[width=2.5in]{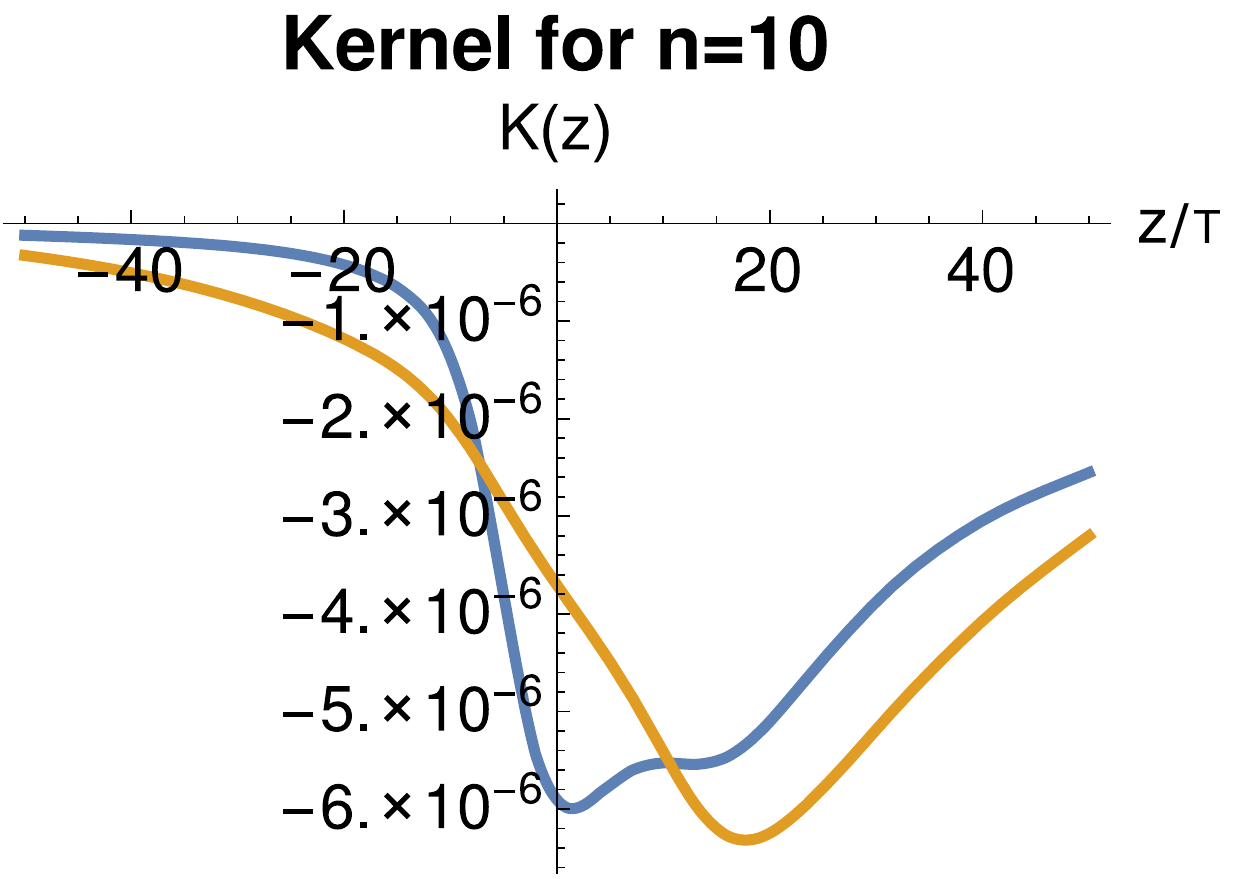}
\includegraphics[width=2.5in]{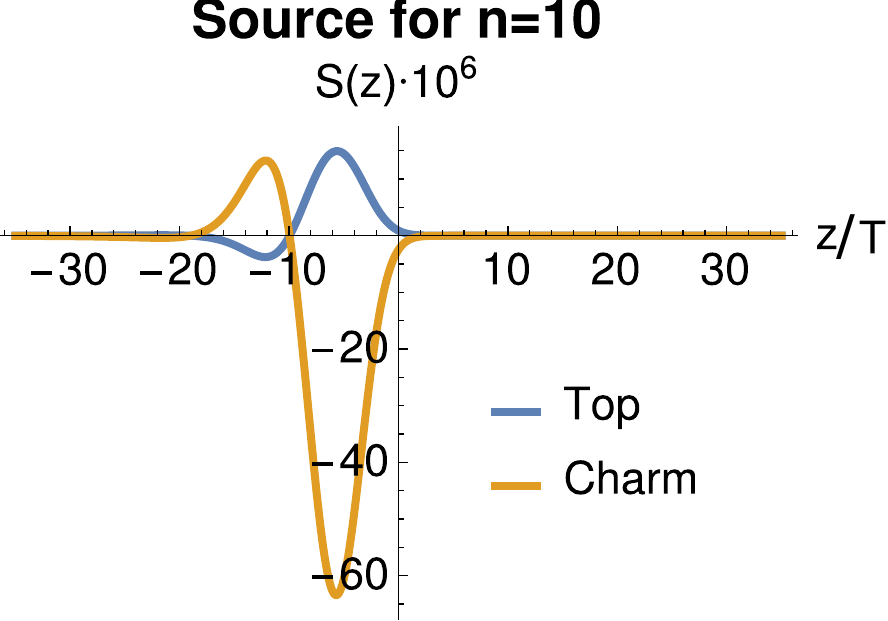}\\
\includegraphics[width=2.5in]{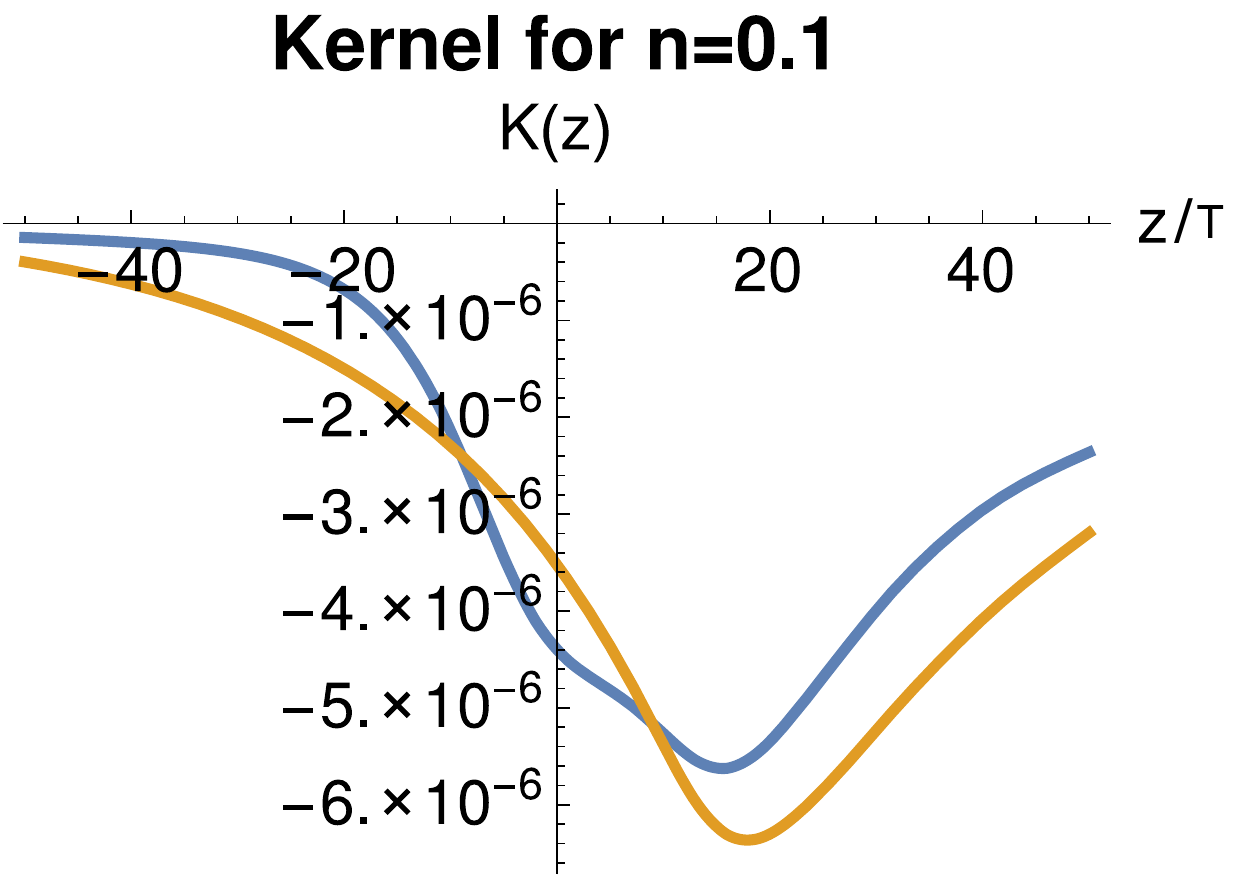}
\includegraphics[width=2.5in]{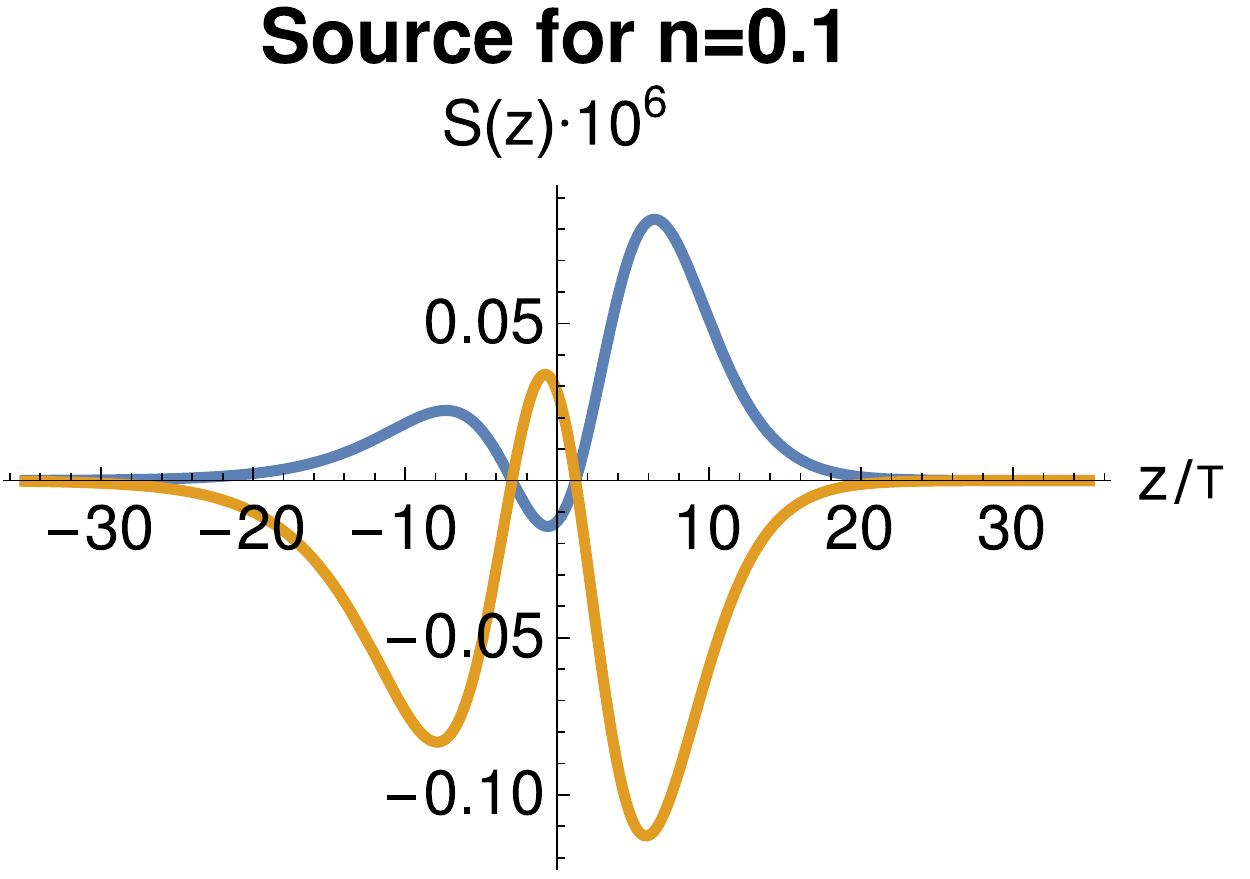}
\caption{
Kernel and source terms for two parametrizations of the Yukawa coupling variation given by Eq.~(\ref{eqn:generalParam}) and (\ref{yukawamatrix}) applied to the (Top,Charm) system only. The symmetric phase corresponds to positive $z$. The kernels are not very dependent on the exact form of the Yukawa couplings as they only enter the calculation through the Yukawa interaction rates. The source term on the other hand gets the expected shift towards the symmetric  phase for small values of $n$ and it also gets a large suppression of the amplitude by $\sim 2$ orders of magnitude from the value of the Higgs VEV. The corresponding baryon asymmetries are: $\eta_B\approx 5.5 \cdot 10^{-10}$ for $n=10$ and $\eta_B\approx 2.1 \cdot 10^{-12}$ for $n=0.1$. 
Figure is taken from \cite{Bruggisser:2017lhc}.
}
\label{fig:GeneralKernelAndSource} 
\end{figure}

To understand what controls the final prediction for the baryon asymmetry, it is enlightening to plot separately those two physical effects $\bar{\cal S}_i(z)$ and  $K_i(z)$, for different assumptions of the source term and bubble configuration, as will be done in Figure \ref{fig:GeneralKernelAndSource}.

The case where only the Top quark Yukawa coupling varies across the bubble wall was  studied 
in ~\cite{Bodeker:2004ws, Espinosa:2011eu}. If during the EW phase transition the top mass exhibits a changing complex phase, $m_t = | m_t(z)| \exp(i \, \theta(z))$, the CP violating  source in the Boltzmann equation is
\begin{equation}
 \Im\left[V^\dagger {m^\dagger}'' m V\right]_{tt}=\Im\left[{m_t^\dagger}'' m_t \right]
=- \left[|m_t|^2 \theta'\right]' 
\end{equation}
which leads to sufficient CP-violation for successful EWBG.
It does not work for other (light) flavours \cite{Bruggisser:2017lhc} if only one of them is varying. On the other hand, when more than one flavour is changing, and provided that Yukawas are of order one in the symmetric phase, sufficient baryon asymmetry can be produced, even with leptons \cite{Bruggisser:2017lhc}.

\begin{figure}[!t]
\centering\includegraphics[width=2.5in]{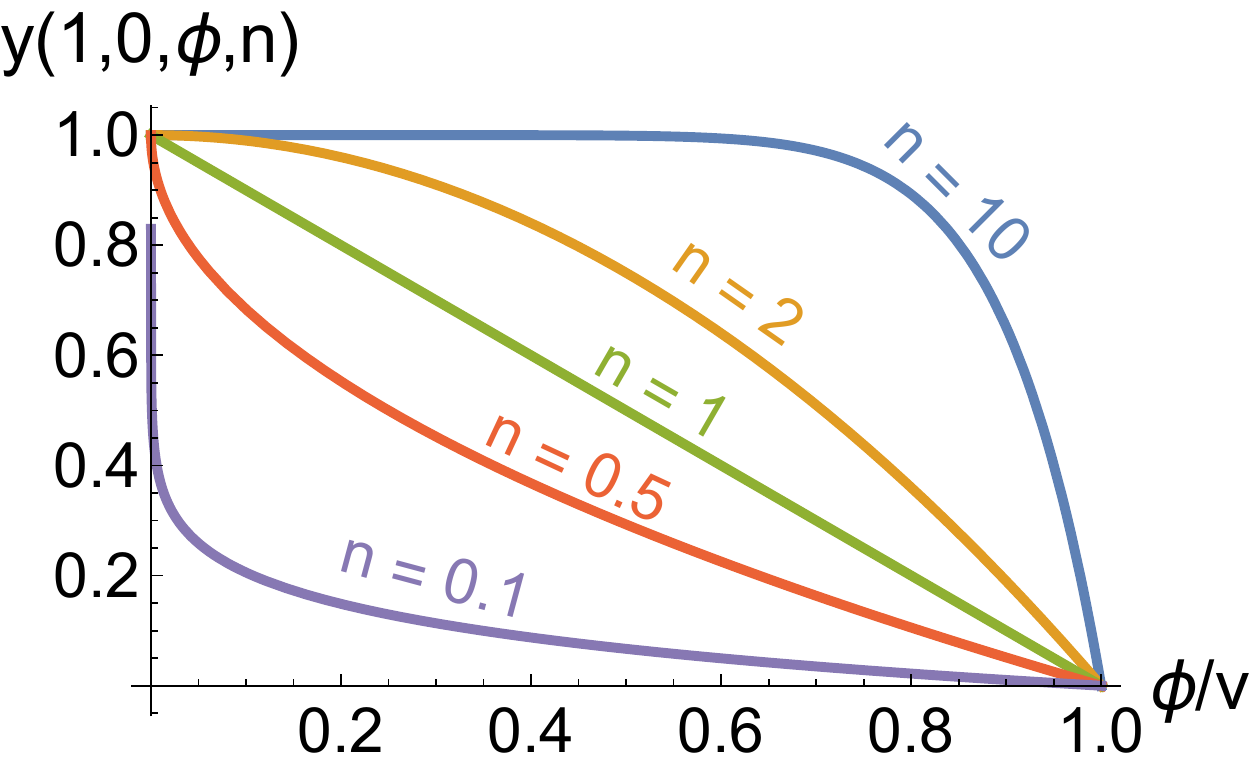}
\caption{Parametrization of the Yukawa variation across the bubble wall given in eq.~\ref{eqn:generalParam} for different values of $n$ for light flavours (it does not apply for the top quark). Figure is taken from \cite{Bruggisser:2017lhc}.}
\label{fig:yukawasGeneral}
\end{figure}

Before studying specific models of varying Yukawas, it is useful to consider some generic parametrization of the Yukawa variation to learn about the requirements on the flavour model  to explain the baryon asymmetry. A similar approach was presented in \cite{Baldes:2016rqn} to study the impact of varying Yukawas on the nature of the EW phase transition. In practise, one should include the degrees of freedom responsible for the dynamics of the Yukawas, typically the flavon field. However, we can  replace the flavon VEV by a function of the Higgs VEV once a field trajectory is assumed during the EW phase transition. 
 As the flavon is coupled to the Higgs, the Yukawas effectively depend on the value of the Higgs VEV. This can be parametrized as follows
\begin{equation}\label{eqn:generalParam}
  y(y_0,y_1,\phi,n)=(y_0-y_1)\left[1-\left(\frac{\phi}{v}\right)^n\right]+y_1
\end{equation}
where $y_0$ ($y_1$) is the value of the coupling in the symmetric (broken) phase, $\phi$ the Higgs VEV, $v$ the Higgs VEV in the broken phase (minimum of the potential) and $n$ is a free parameter that indicates where the change takes place. For a large value of $n$ the Yukawa coupling keeps the value of the symmetric phase until deep inside the broken phase, whereas for small values of $n$ the Yukawa changes rapidly in front of the wall (i.e. in the symmetric phase). This is illustrated on fig.~\ref{fig:yukawasGeneral} where the Yukawa coupling is shown as a function of $\phi$ for different choices of $n$. The choice of $n$ is a parametrisation of the trajectory in the (Higgs, Flavon) field space during the phase transition, which can be determined through a dedicated analysis of tunneling in a 2-field space \cite{Bruggisser:2018mus,Bruggisser:2018mrt,Baldes:2018nel}. Typically, for a simple polynomial potential, the lighter the flavon, the larger $n$ can be. On the other hand, for a flavon much heavier than the Higgs, tunnelling will tend to happen in  the flavon direction and therefore this will correspond to a small $n$ value. In this case, EW symmetry breaking happens after most of the flavour structure has already emerged, which will suppress CP-violation from Yukawa variation during the EW phase transition and therefore when sphalerons get out-of-equlibrium.
The precise form of the  ansatz  (\ref{eqn:generalParam}) is not fundamental for the argument we want to make. In fact it does not capture all behaviours encountered in specific models. 
However the only crucial feature is that the Yukawa couplings are of order one in the symmetric phase  and remain large in some fraction of the bubble wall. 

As the CP violating source (\ref{eq:CPsource}) is proportional to the second derivative of the mass parameter we expect the source for small (big) $n$ to be localized outside (inside) the broken phase. It would therefore seem beneficiary to have a model which would map onto a parametrization with a small $n$. As the weak sphaleron is most active in the symmetric phase, the baryon asymmetry can be augmented by creating the CP asymmetry in front of the bubble wall and therefore evading the suppression from the diffusion through the bubble wall. However, the CP violating source is also proportional to the mass parameter and hence every source deep inside the symmetric phase gets suppressed by the Higgs VEV which is rapidly approaching zero in the symmetric phase. This is shown on figure~\ref{fig:GeneralKernelAndSource} focussing on the (top, Charm) system where we report the results for the Yukawa matrix:
\begin{equation}
	Y=\left(\begin{array}{cc}
		e^{i \theta}y(1,0.008,\phi,n) & y(1,0.04,\phi,n) \\
		y(1,0.2,\phi,n) & y(1,1,\phi,n)
	\end{array}
	\right) \qquad \text{with} \qquad \theta=1\, .
	\label{yukawamatrix}
\end{equation}
We see a two orders of magnitude decrease in the amplitude of the source term. This cannot be compensated by the change in the kernel when going from the broken phase to the symmetric one,  which is less than one order of magnitude.
There is even an additional suppression in the case of a source term that is located in the symmetric phase. The source term for the particle $i$ is proportional to $\text{Im}\left[V^\dagger {m^\dagger}'' m V\right]_{ii}$. It is easy to show that the matrix $\text{Im}\left[V^\dagger {m^\dagger}'' m V\right]$ is traceless for constant complex phases in the Yukawa couplings. Therefore the source for the top always has the opposite sign of the source for the charm. If this were the full source term there would be an almost perfect (up to differences in the kernels) cancellation between the contribution from the top and the charm. The full source  also  depends on the mass of the particle considered. For a source that is localized inside the broken phase, where the particles have very different masses,  a big cancellation between the two contributions is avoided. On the other hand, for a source located in the symmetric phase, where both particles have a nearly vanishing mass, the cancellation is very much unsuppressed. 

This analysis shows that for the parametrisation (\ref{eqn:generalParam}) and assuming the same $n$ for all flavours, models with a source located inside the broken phase are generally more effective in producing the baryon asymmetry even though we naively expect the mechanism to be more effective when the source is located in front of the bubble wall. However, when studying Randall--Sundrum models focussing on the (Top, Charm) system (which would then correspond to profiles with different $n$), we find that it is still possible to create a sizable baryon asymmetry using sources that are located inside the symmetric phase \cite{Bruggisser:2017lhc}.

\section{Effect of varying Yukawas on the nature of the electroweak phase transition}

Varying Yukawas during EW symmetry breaking have an impact on the nature of the EW phase transition \cite{Baldes:2016rqn,Baldes:2018nel}. The first effect comes from the $T=0$ one-loop Higgs potential. Large Yukawas in the symmetric phase can lead to a significant decrease of the potential in the region $0 < \phi < v$. This can weaken the phase transition. More interesting is
the $T\neq 0$ one-loop contribution from the fermions, 
   $V^{T}_{f}(\phi,T)=-\frac{g_fT^{4}}{2\pi^{2}}J_{f}\left(\frac{ m_{f}(\phi)^{2} }{ T^{2} } \right),
$
where $g_f$ is the number of internal degrees of freedom and $J_{f}(x^{2})$ has a high-temperature expansion for $x^{2} \ll 1$, $        J_{f}(x^{2}) \approx \frac{7\pi^4}{360}-\frac{\pi^{2}}{24}x^{2}-\frac{x^{4}}{32}\mathrm{Log}\left[\frac{x^{2}}{13.9}\right]$. For decreasing Yukawas,  the second term leads to a barrier between the symmetric and broken phases,
$	\delta V \equiv V^{T}_{f}(\phi,T)-V^{T}_{f}(0,T)\approx \frac{gT^{2}\phi^{2}[y(\phi)]^{2}}{96}.$
This leads to a cubic term in $\phi$, e.g. for $y(\phi) = y_{1}(1-\phi/v)$:	$
	\delta V \approx \frac{g y_{1}^{2} \phi^{2} T^{2}}{96} \left(1-2\frac{\phi}{v}+\frac{\phi^{2}}{v^{2}}\right)$.
This can result in a first order phase transition.
 Finally, large Yukawas at $\phi \sim 0$ significantly increase the Higgs thermal mass, 
$
   \Pi_{\phi}(\phi,T)  =  \left(\frac{3}{16}g_{2}^{2}+\frac{1}{16}g_{Y}^{2}+\frac{\lambda}{2}+\frac{y_{t}^{2}}{4}+\frac{gy(\phi)^2}{48}\right)T^{2}.
      $
      The novelty is  the dependence of the thermal mass on $\phi$, which comes from 
the $\phi$-dependent Yukawa couplings (these do not enter into the thermal masses for the $W$ and $Z$ bosons at this order). The effect of this term is to lower the effective potential at $\phi=0$, with respect to the broken phase minimum, as long as $\Pi_{\phi}(0,T_{c}) \gg \Pi_{\phi}(\phi_{c},T_{c})$. By lowering the potential at $\phi=0$, the phase transition is delayed and strengthened, 
 which, through the Daisy resummation, lowers the potential close to the origin $\phi\sim0$, delaying the phase transition and thereby increasing $\phi_c/T_c$. All these effects favour  a strong first-order EW phase transition. Besides, the 1-loop thermal effects from additional heavy fermions with flavon-dependent masses present in many flavour models may impact the phase transition. Finally, a dominant effect may be due to tree level barriers generated by couplings between the Higgs and the flavon.

After having shown the interesting opportunities for EWBG 
offered by varying Yukawas during the EW phase transition, we now discuss implementations in two popular classes of models of flavor hierarchies,  Froggatt--Nielsen models and  Randall-Sundrum  models respectively. 

\section{Low-scale Froggatt-Nielsen models}

The Froggatt--Nielsen (FN) mechanism was proposed as a possible explanation for the mysterious hierarchy in the observed fermion masses~\cite{Froggatt:1978nt}. While there is a plethora of  implementations of this mechanism in the literature, the question of the dynamics and cosmology of this mechanism has not been addressed. In \cite{Baldes:2016gaf}, we studied whether this dynamics could viably happen at the EW scale (while it is typically assumed that this  mechanism occurs at much higher scales). 

\begin{figure}[b]
\centering\includegraphics[width=2.5in]{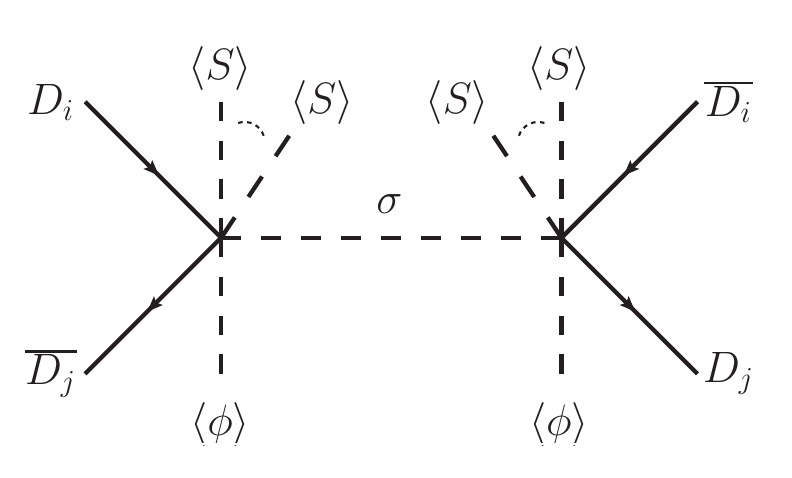}
\caption{  Diagram contributing to neutral meson oscillations in Froggatt Nielsen models mediated by flavon exchange $\sigma$. The number of $\langle S \rangle$ insertions depends on the Froggatt-Nielsen charge of the quarks. Each VEV insertion leads to a suppression of order $\epsilon_s=\langle S \rangle/\Lambda_s$. Figure is taken from \cite{Baldes:2016gaf}.}
\label{fig:mesons}
\end{figure}

The main idea of the FN mechanism is that Standard Model (SM) fermions carry different charges  under a new flavor symmetry, which is spontaneously broken by the VEV  $\langle S \rangle$ of a new scalar field, the so-called flavon field $\sigma$.  
Together with the flavon, a relatively large number of vector-like quarks are introduced to communicate the breaking of the flavor symmetry. The flavon field as well as all new quarks carry a Froggatt-Nielsen charge. The Yukawa couplings of the SM quarks are then generated by chain-like tree-level diagrams in which the difference in FN charge of the SM quarks is saturated by a number of insertions of the flavon field. Once the flavon obtains a VEV, the Yukawa couplings of the SM quarks are generated and are of order $(Y \langle S \rangle /M)^m$ where $m$ is the difference in FN charge, $Y$ is an order one coupling between the new quarks and the flavon and $M$ is the typical mass of a vector-like quark.
We define the scale of new physics as $\Lambda_s = M/Y$
and we introduce the ratio
\begin{equation} 
	\label{eq:epsdefined}
	\epsilon_{s} \equiv \frac{ \langle S \rangle }{ \Lambda_{s} }
		\end{equation}
The hierarchy in the SM Yukawa couplings is then given by the different powers of $\epsilon_{s}$. 
For example, for some appropriate choice of charge assignments,  one finds the effective quark Yukawa couplings to be 	$
	y^{t}_{\rm{eff}} \sim 1, \ y^{c}_{\rm{eff}} \sim \epsilon_{s}^{3}, \ y^{u}_{\rm{eff}} \sim \epsilon_{s}^{7}, 
	y^{b}_{\rm{eff}} \sim \epsilon_{s}^{2}, \  y^{s}_{\rm{eff}} \sim \epsilon_{s}^{4}, \  y^{d}_{\rm{eff}} \sim \epsilon_{s}^{6}$. Similarly one finds the pattern of CKM matrix elements
	$	|V_{us}| \sim |V_{cd}| \sim \epsilon_{s}, \ |V_{cb}| \sim |V_{ts}| \sim \epsilon_{s}^{2}, \ |V_{ub}| \sim |V_{td}| \sim \epsilon_{s}^{3}$.
The observed quark masses and mixing can therefore be accommodated with 
\begin{equation}
\epsilon_{s} \sim 0.2  \ \mbox { today}.
\end{equation}
The explanation of the fermion mass hierarchy in the FN mechanism does not depend on the value of $\Lambda_s$, only on the ratio $\epsilon_s=\langle S \rangle/\Lambda_s$. If $\epsilon_s$ was of order 1 rather than 0.2 we would not observe any hierarchical structure and all Yukawas would be ${\cal O}(1)$. We are interested in the possibility that
\begin{equation}
\epsilon_s \sim {\cal O}(1) \  \mbox{ before the EW phase transition,}
\end{equation}
 as motivated by EWBG. As discussed above, this requirement imposes the flavon to be not much heavier than the Higgs (to vary at the same time as the Higgs) and therefore $\Lambda_s$  to be not much higher than the TeV scale. On the other hand, 
there are strong experimental constraints on $\Lambda_s$ from meson oscillations as illustrated in Figure \ref{fig:mesons}, typically of the type
\begin{equation}
m_{\sigma} \Lambda_s \gtrsim (\mbox{a few TeV})^2 \, .
\end{equation}
The conclusion of \cite{Baldes:2016gaf}  is that implementing Yukawa coupling variation during the EW phase transition in un-tuned Froggatt-Nielsen models is not possible without violating experimental bounds.
A way to circumvent this conclusion was recently advocated by making the EW phase transition occurring at a temperature much higher than today's EW scale, through a symmetry-non-restoration effect~\cite{Baldes:2018nel}. We now move to another promising implementation.

\section{Naturally varying Yukawas in Randall-Sundrum model}

One of the very attractive features of the Randall--Sundrum  (RS) model \cite{Randall:1999ee}
is that in addition to bringing a new solution to the Planck scale/weak scale hierarchy problem, it offers a new tool to understand flavour and explain the hierarchy of fermion masses \cite{Gherghetta:2000qt,Huber:2000ie,Gherghetta:2010cj}. The setup is a slice of 5D Anti-de Sitter space (AdS$_5$) which is bounded by two branes, the UV (Planck) brane where the graviton is peaked, and the IR (TeV) brane hosting  the Higgs (which therefore does not feel the UV cutoff). Fermions and gauge bosons are free to propagate in the bulk.
In this framework, the effective 4D Yukawas of SM fermions are given by the overlap of their 5D wavefunctions with the Higgs. 
Since the Higgs is localised towards the IR brane to address the Planck scale/weak scale hierarchy, 
small Yukawas are achieved if the fermions live towards the UV brane so that the overlap between the fermions and the Higgs is suppressed. On the other hand, heavy fermions such as the top quark are localised near the IR brane. This setup leads to a protection from large flavour and $CP$-violation via the so-called RS-GIM mechanism.

The key feature for flavour physics is therefore the localisation of the fermions in the AdS$_5$ slice, which determines the effective scale of  higher-dimensional  flavour-violating operators.
The profile of a fermion is  determined by its 5D  bulk mass. 
Because of the AdS$_5$ geometry, modifications of order  one in the 5D bulk mass have a substantial  impact on the fermionic profile and therefore on the effective 4D Yukawa coupling. In fact, the 4D Yukawa couplings depend exponentially on the bulk mass parameter.
Randall-Sundrum models are holographic duals of 4D strongly coupled theories. In this picture, the Higgs is part of the composite sector. 
The size of the Yukawa couplings is then determined by the degree of compositeness of the states that are identified with the SM fermions.
Indeed fermions localised near the UV brane are dual to mainly elementary states leading to small Yukawas while fermions localised near the IR brane map to mainly composite states with correspondingly large Yukawas.

In the usual picture, the bulk mass parameter is assumed to be constant. On the other hand, it is quite well motivated to consider that this bulk mass is dynamical and generated by coupling the fermions to a bulk scalar field which in turn obtains a VEV. 
We can then expect a position-dependent bulk mass as this VEV is generically not constant along the extra dimension.
In fact, the simplest mechanism for radion stabilisation, due to Goldberger and Wise \cite{Goldberger:1999uk}, consists in introducing a bulk scalar field which obtains a VEV from potentials on the two branes.
The most minimal scenario to dynamically generate the bulk mass is then to use this bulk scalar.
Interestingly, during the process of radion stabilisation, the profile of the Goldberger-Wise scalar VEV changes. When the latter is coupled to the fermions, this induces a change in the bulk masses of the fermions which in turn affects their wavefunction overlap with the Higgs on the IR brane and thus the Yukawa couplings. 
The RS model with bulk fermions therefore naturally allows for a scenario of varying Yukawa couplings during the EW phase transition.
The cosmological dynamics of Yukawa couplings in this context was studied in  \cite{vonHarling:2016vhf}.

The emergence of the EW scale in RS models comes during the stabilisation of the size of the AdS$_5$ slice. 
At high temperatures, the thermal plasma deforms the geometry and the IR brane is replaced by a black hole horizon. 
Going to lower temperatures, eventually a phase transition takes place and the IR brane emerges. This phase transition is typically strongly first-order and proceeds via bubble nucleation. 
The walls of these bubbles then interpolate between AdS$_5$ with an IR brane at infinity and at a finite distance. 
In the dual 4D theory, this transition is described by the dilaton $\sigma$ -- which maps to the radion -- acquiring a VEV. 
In the effective 4D lagrangian for the Higgs, a VEV for the radion spontaneously breaks conformal invariance and corresponds to adding a mass term for the Higgs which induces EW symmetry breaking. To show that the Higgs and dilaton get a VEV at the same time during the phase transition requires a precise analysis of tunnelling in the 2-field potential, as done in Ref.~\cite{Bruggisser:2018mus,Bruggisser:2018mrt}.
To realise a model where the Yukawas are larger during the phase transition (as needed if we want to use the SM Yukawas as the unique $CP$-violating source during EWBG \cite{Bruggisser:2017lhc,Berkooz:2004kx}),  we ask for the Yukawas to become larger when the IR brane is pushed to infinity.
For a constant bulk mass term $c \ k$ where $k \sim {\cal O}(M_{Pl})$ is the AdS$_5$ curvature scale
\begin{equation} 
S \supset \int d^5x \sqrt{g} \ c \ k \ \bar{\psi} \psi
\end{equation}
the effective 4D Yukawa coupling between the SM fermions and the Higgs is given by 
\begin{equation} 
Y(\sigma)= \lambda \sqrt{\frac{1-2c_L}{1-\sigma^{1-2c_L}}} \sqrt{\frac{1-2c_R}{1-\sigma^{1-2c_R}}}
\label{eq:effYuk}
\end{equation}
where $\sigma= k \ e^{-ky}$ is the radion/dilaton, $y$ is the interbrane distance (i.e. position of the IR brane as UV brane is at $y=0$) and $\lambda$ is the  5D Yukawa coupling.  
The bulk mass parameters $c_L$ and $ c_R$ correspond to the 5D left-handed $SU(2)$ doublet and the right-handed singlet vector-like fermions respectively, of a given generation.
For $c_L,c_R> 1/2$, Eq.~(\ref{eq:effYuk}) becomes exponentially suppressed. This shows how large hierarchies between the 4D Yukawa couplings can be obtained in RS starting from bulk mass parameters and 5D Yukawa couplings of order one in units of the AdS scale $k$. In this setup the Yukawa couplings depend on $\sigma$ thus on the position of the IR brane. Since the light quarks are all localized towards the UV brane, however, their Yukawa couplings decrease when the IR brane is sent to infinity, $\sigma \rightarrow 0$. Correspondingly, they are small in a large portion of the bubble wall during the phase transition and $CP$-violation is suppressed. 

\begin{figure}[t]
\centering\includegraphics[width=5.25in]{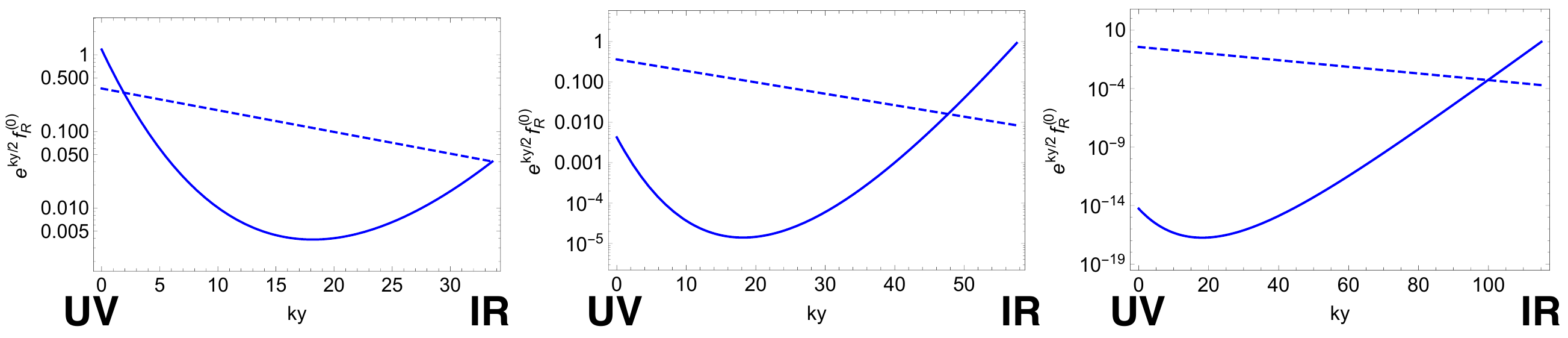}
\caption{  From left to right, the IR brane is being pushed away from the UV brane with the hierarchies $\sigma = 2.5 \times 10^{-15}, 10^{-25}$ and $10^{-50}$ respectively. 
Plotted is the normalized wavefunction of the right-handed charm quark along the extra dimension.
The solid curve is the wavefunction for the position-dependent bulk mass, whereas the dashed curve is for the usual case with constant bulk mass. Figure is taken from \cite{vonHarling:2016vhf}.}
\label{fig:backintime}
\end{figure}

Now if we assume instead that the bulk fermion mass term comes from a Yukawa coupling with the Goldberger-Wise scalar $ \phi_{GW}$ i.e. 
\begin{equation} 
c  \, k \, \overline{\psi } \psi \, \rightarrow  \, \rho \, \phi_{GW} \, \overline{\psi } \psi \, ,
\end{equation}
where $\rho$ is a constant of dimension -1/2, then the fermion mass terms  become position-dependent and decrease when the IR brane is pushed to large y and $\sigma \to 0$. 
Since smaller bulk masses make the wavefunctions grow faster towards the IR, this leads to fermions which become increasingly IR-localized when the IR brane is pushed to infinity. The wavefunction overlap with the Higgs near the IR brane then grows. 
As a result, the 4D effective Yukawa coupling increases when the IR brane is sent to infinity \cite{vonHarling:2016vhf}, as shown in Figure \ref{fig:backintime}.
Such behaviour can be understood from eq. (\ref{eq:effYuk}) by using $\sigma$-dependent values of $c_L$,  $c_R$ which decrease when $\sigma \to 0$. 
This mechanism can be relevant for $CP$-violation for all quarks and enables a large variation of the Yukawa couplings, from values of order one to today's small values of the light quarks. 

\begin{figure}[t]
\centering\includegraphics[width=1.6in]{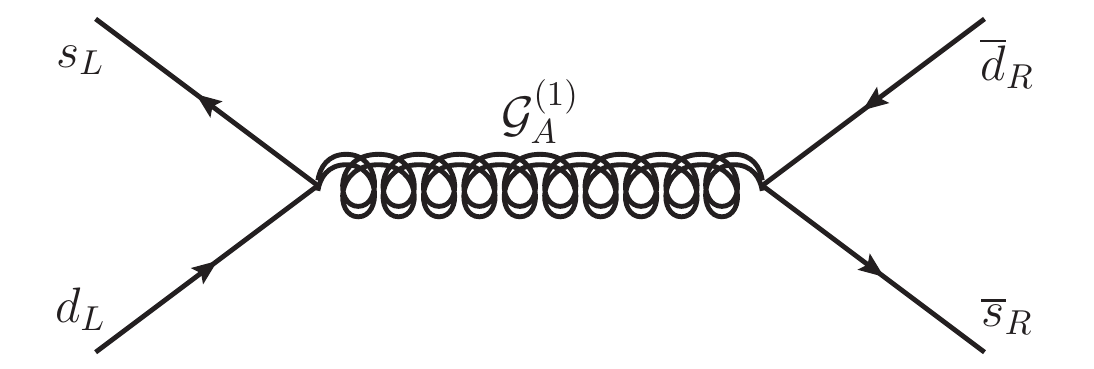}
\centering\includegraphics[width=1.8in]{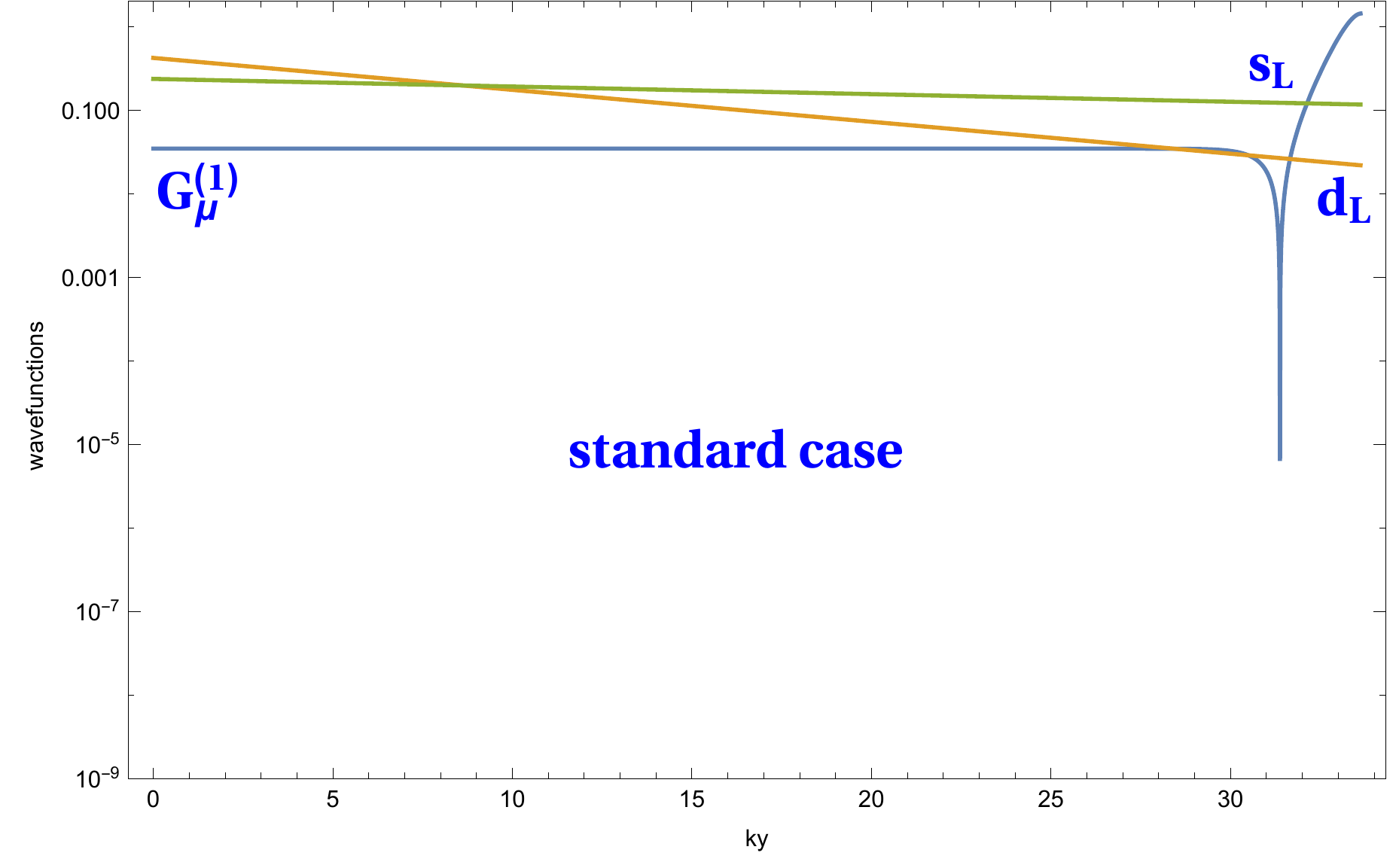}
\centering\includegraphics[width=1.8in]{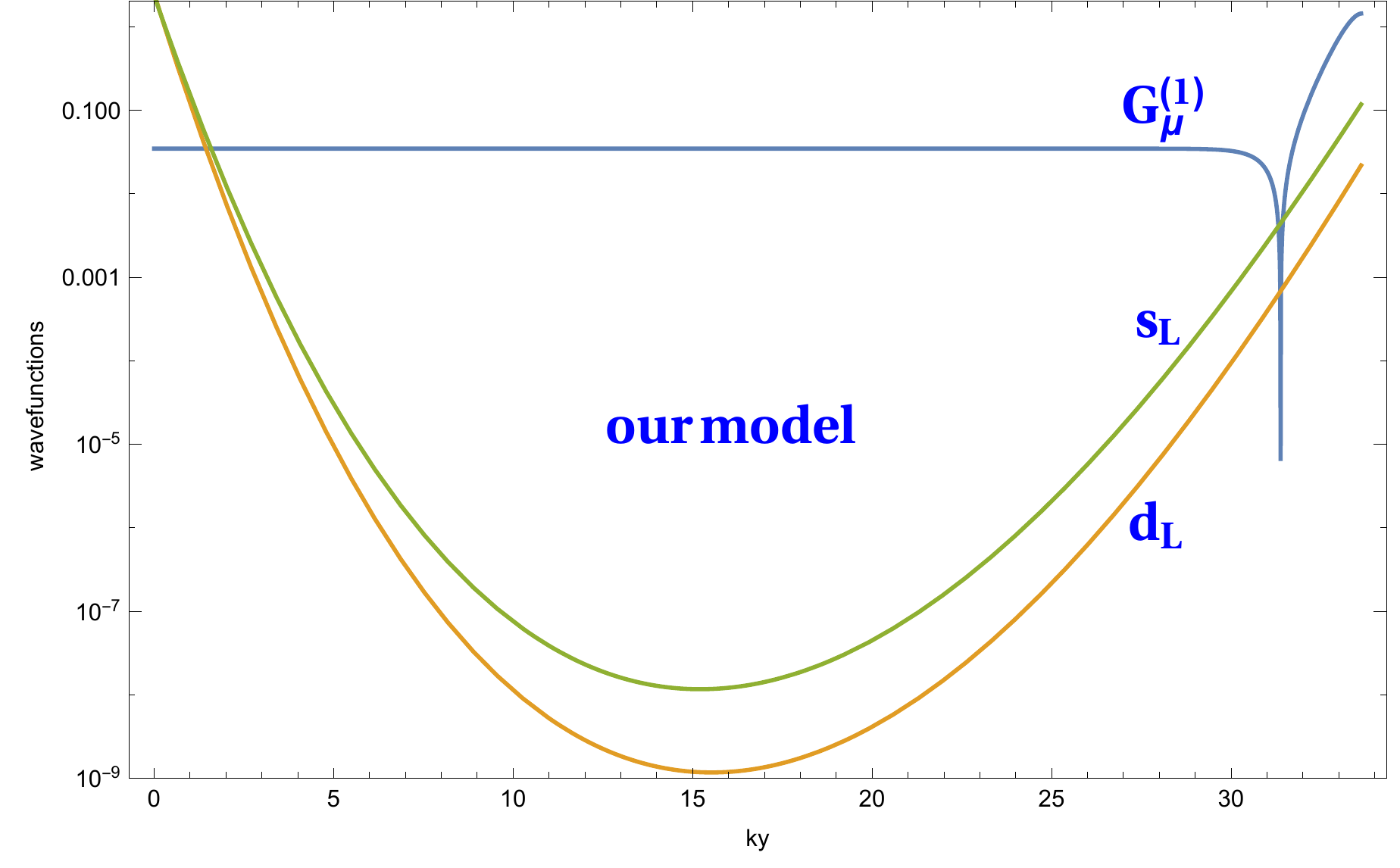}
\caption{ (a) The most important new contribution to  $CP$-violation in $K$-$\overline{K}$-mixing depends on the coupling between the KK gluon and SM $b$ and $d$ quarks which depends on the overlap integral of the three wave functions. (b) Profiles of SM quarks and KK gluon in standard RS with constant bulk fermion mass term. (c) Profiles of SM quarks and KK gluon in RS with bulk fermion mass term given by VEV of Goldberger-Wise field.}
\label{fig:overlap}
\end{figure}

A particularly interesting spinoff in this construction is that modified fermion wave functions give suppression of CP-violating processes which are on the other hand very constraining in usual RS. 
In fact, the overlap between the wave functions of the KK gluon and the light fermions is now reduced in the vicinity of the IR brane, while remaining the same on the IR brane itself -- where the Higgs is localized -- in order to lead to the same 4D Yukawas. The decreased overall overlap causes the 4D effective coupling of light quarks to KK gluons to become suppressed, which eases constraints
in $K-\bar{K}$ mixing as illustrated in Figure \ref{fig:overlap}. The resulting bound on the KK scale is weakened by a factor of 3, significantly ameliorating the little hierarchy problem in RS \cite{vonHarling:2016vhf}.

Another realisation  mentioned in \cite{vonHarling:2016vhf}  to induce varying Yukawas is to add an operator on the IR brane that effectively changes the value of the Yukawa coupling as the position of the IR brane changes.
This mechanism enables variations of order one for the Yukawas and can be relevant for $CP$-violation if applied to the top quark.

We conclude with a comparison of  Froggatt-Nielsen and Randall-Sundrum models. In both cases, the coupling of the new scalar  (flavon in FN and radion/dilaton in RS) to SM fermions is proportional to the Yukawa coupling so there is a suppression for light quarks. However, this is still not sufficient for FN models where there is a $1/\epsilon_s$ enhancement factor of the flavon-SM fermion coupling due to its charge under the flavour symmetry. The RS construction on the other hand appears as a viable natural framework of Yukawa variation during the EW phase transition. 
A purely 4D description in the framework of composite Higgs models 
is presented in \cite{Bruggisser:2018mus,Bruggisser:2018mrt}.
One intriguing way to reconcile Yukawa coupling variation during the EW phase transition with Froggatt-Nielsen models is if the EW phase transition happens at high scale, see Ref.~\cite{Baldes:2018nel}.


\section{Conclusion}

Time-dependent CP-violating sources can make electroweak baryogenesis compatible with Electric Dipole Moment constraints and are well-motivated. Strong CP violation from the QCD axion is an example \cite{Servant:2014bla}. In this article we focused on Weak CP violation. The dynamical interplay between the physics at the origin of the flavour structure and electroweak symmetry breaking is an intriguing possibility. 
The specific experimental test of this weak scale flavour cosmology is a key question and a non-trivial one that deserves further study. It is also very model-dependent. Signatures and phenomenological implications were discussed in the case of Froggatt-Nielsen models in \cite{Baldes:2016gaf}. 
In Randall-Sundrum and composite Higgs models, one has to search for the radion/dilaton field  and measure its couplings at the LHC and future colliders~\cite{Bruggisser:2018mus,Bruggisser:2018mrt}.  Heavy fermionic resonances are also typically predicted in all studied benchmark models.
More generally, a stochastic background gravitational waves at LISA is a generic signature of a strongly-first-order electroweak phase transition \cite{Caprini:2015zlo}.


\enlargethispage{20pt}






\ack{I am grateful to Iason Baldes, Sebastian Bruggisser, Benedict Von Harling, Alexey Matsedonskyi and Thomas Konstandin for fruitful collaborations.}



\end{document}